\documentclass[aps,prep,preprintnumbers,amsmath,amssymb,nofootinbib]{revtex4}
\usepackage{amssymb}
\usepackage{lscape}
\usepackage{amsmath}
\usepackage{rotating}
\usepackage{bm}
\usepackage{graphicx}
\usepackage{epsfig}
\begin{document}
\title{Discrete Modes in Gravitational Waves from the Big-Bang}
\author{$^{2}$ Luis Santiago Ridao,  $^{1,2}$ Mauricio Bellini
\footnote{E-mail address: mbellini@mdp.edu.ar} }
\address{$^1$ Departamento de F\'isica, Facultad de Ciencias Exactas y
Naturales, Universidad Nacional de Mar del Plata, Funes 3350, C.P.
7600, Mar del Plata, Argentina.\\
$^2$ Instituto de Investigaciones F\'{\i}sicas de Mar del Plata (IFIMAR), \\
Consejo Nacional de Investigaciones Cient\'ificas y T\'ecnicas
(CONICET), Argentina.}

\begin{abstract}
We develop a new approach to gravitational waves in which the Einstein equations are governed by the cosmological constant which is related to
the existence of a manifold which is closed. We study an example in which
the matter Lagrangian is described by the scalar (inflaton) field. There are only three
dynamical solutions. In one of them the universe is initially static but
begins to increase until an inflationary stage. We calculate the dynamics of GW in this primordial pre-inflationary stage of the universe.
We found that there should be an infinite number of polarization modes in order to the fields can be quantized.
Finally, we calculate the energy density due to the gravitational waves.
\end{abstract}
\maketitle

\section{Introduction}

When one tries to describe the early universe, GR is applied beyond its domain of
validity. The quantum effects which dominate in this epoch are
expected to resolve the singularity. In particular, the existence
of the cosmological singularity in the framework of Loop Quantum
Gravity (LQG)\cite*{4,4a} has been subject of study in the last years.
However, at the present time, it is not possible to realize a consistent quantum gravity theory which leads to the unification of gravitation with the other forces.
In particular, the theory of gravitational waves (GW) is a rich subject that brings together different domains such as general relativity, field theory, astrophysics and cosmology.
At present various gravitational-wave detectors, after decades of developments, have reached a sensitivity where there are significant chances of detection, and future improvements are expected to lead, in a few years, to advanced detectors with even better sensitivities \cite*{pl3}.
There are good reasons to expect that the Universe
is permeated also by a stochastic background of GW generated in the early universe. In particular, the fossil
GW becoming from the big-bang should be a great success if it were detected. Recently
has been reported GW produced during the inflationary epoch\cite*{BICEP2} which are compatible
with a energy scale of $10^{16} \,GeV$. However, more data are required to confirm
the above situation. During the inflationary expansion, the universe suffered
an exponential accelerated expansion driven by a scalar (inflaton)
field with an equation of state close to a vacuum dominated
one\cite*{ua,ua1,LB}. The most conservative assumption is that the
energy density $\rho =P/\omega $ is due to a cosmological
parameter which is constant and the equation of state is given by
a constant $\omega =-1$, describing a vacuum dominated universe
with pressure $P$ and energy density $\rho$. Inflationary cosmology
can be recovered from a 5D vacuum\cite*{nos,nosb,nosc}, and is very
consistent with current observations
of the temperature anisotropy of the Cosmic Microwave Background
(CMB)\cite*{5}. The most popular model of supercooled inflation is
chaotic inflation\cite*{6}, but there are many models which are good candidates.
In this model the expansion of the
universe is driven by a single scalar field called inflaton. At
some initial epoch, presumably the Planck scale, the scalar field
is roughly homogeneous and dominates the energy density, which
remains almost constant during all the inflationary epoch.
It is well known that the
inflationary cosmology also generates a
background of gravitational waves \cite*{GW1}. Dark energy
cosmological scenarios have been intensively studied in the last
years\cite*{BCNO}. The scenarios there described can explain the
generation of gravitational waves on cosmological, but not on
astrophysical scales.

During inflation the energy scales are of the order of $(10^{16}-10^{18})\,\, GeV$, but it is expected that can exist an earlier epoch of the universe in which it begins
to increase until becoming in a de Sitter expansion. In this epoch the energy scales in the universe should be bigger, of the order of $10^{19}\,\, GeV$, which is the Planckian order of energy. Such epoch is called Pre-inflation and is necesary to explain the accelerated expansion of the universe known as inflation, which is very important because solves many cosmological problems such as, for example, can explain the flatness, isotropy and homogeneity of the present day universe such as can explain the nonexistence of magnetic monopoles. Of course all these predictions are only valid on cosmological scales which actually are in the range $(10^8-10^{10})\,\,l.y.$. Inflation is also very important because is capable to explain the genesis of primordial structure formation in the universe, which is later makes possible the galactic formation. When the inhomogeneities of the scale factor which driven inflation are unstable their modes cross the causal horizon which is related to the cosmological constant $\Lambda$. After inflation, when matter acquire mass these modes re-enter the horizon and become causally connected again. During inflation the size of the horizon was of the order of $10^{5}\,\, \sqrt{G}$\footnote{The value of the gravitational constant is $G \simeq 10^{-38}\, (GeV)^{-2}$.}, but at present is really very small: $\lambda_H \sim 10^{-61}\,\,\sqrt{G}$.

In this work we shall develop a nonperturbative formalism to study gravitational waves in a curved space-time.
After it we shall work an application to the early big-bang universe to study the pre-inflationary dynamics of the
universe. Finally, we shall explore the emission of GW from the early pre-inflationary universe, in order to obtain its
spectral density.

\section{Formalism of Gravitational Waves}

In this section we shall revisite the formalism of GW to obtain the wave equations in a consistent manner.

\subsection{Variation of the matter action}

We consider the general action ${\cal I}$ which describes gravitation and matter
\begin{equation}\label{act}
{\cal I} =\int_V d^4x \,\sqrt{-g} \left[ \frac{R}{2\kappa} + {\cal L}_m\right],
\end{equation}
where $g$ is the determinant of the covariant background tensor metric $g_{\mu\nu}$, $R=g^{\mu\nu} R_{\mu\nu}$ is the scalar curvature, $R^{\alpha}_{\mu\nu\alpha}=R_{\mu\nu}$ is
the covariant Ricci tensor and ${\cal L}_m$ is an arbitrary Lagrangian density which describes matter\footnote{In this paper we shall consider some Lagrangian density related to a metric tensor which is symmetric and free of nonmetricity.}. If we consider an orthogonal base, the curvature tensor will be given written in terms of the Levi-Civita connections
\begin{equation}
R^{\alpha}_{\,\,\,\beta\gamma\delta} = \Gamma^{\alpha}_{\,\,\,\beta\delta,\gamma} -  \Gamma^{\alpha}_{\,\,\,\beta\gamma,\delta} + \Gamma^{\epsilon}_{\,\,\,\beta\delta} \Gamma^{\alpha}_{\,\,\,\epsilon\gamma} - \Gamma^{\epsilon}_{\,\,\,\beta\gamma} \Gamma^{\alpha}_{\,\,\,\epsilon\delta}.
\end{equation}
The variation of the action matter will be
\begin{eqnarray}
\delta  \left[\sqrt{-g} {\cal L}_m \left(g^{\mu\nu},g^{\mu\nu}_{\,\,\,\,\lambda}\right) \right]& = &\sqrt{-g} \left\{ \left[ \frac{\delta {\cal L}_m}{\delta g^{\mu\nu}}- \frac{1}{2} g_{\mu\nu} {\cal L}_m\right] \delta g^{\mu\nu} \right. \nonumber \\
&+& \left.
\left[ \frac{\delta {\cal L}_m}{\delta g^{\mu\nu}_{\,\,\,\,\,,\lambda}}\right] \delta g^{\mu\nu}_{\,\,\,\,\,,\lambda}  \right\}.
\end{eqnarray}
By using the fact that $\delta g^{\mu\nu}_{\,\,\,\,\,,\lambda} = \delta g^{\gamma\mu} g^{\nu\beta} g_{\gamma\lambda,\beta}$, we obtain that
\begin{eqnarray}
\delta\left[\sqrt{-g} {\cal L}_m \left(g^{\mu\nu},g^{\mu\nu}_{\,\,\,\,\,,\lambda} \right)\right] &=&  \sqrt{-g} \, \delta g^{\mu\nu}  \nonumber \\
&\times & \left\{ \frac{1}{2} T_{\mu\nu} +
\left[ \frac{\delta {\cal L}_m}{\delta g^{\mu\nu}_{\,\,\,\,\,,\lambda}}\right] \delta^{\beta}_{\,\,\,\lambda,\beta}\right\}, \nonumber \\
\end{eqnarray}
where $\delta^{\beta}_{\lambda,\beta}=0$, so that
\begin{equation}
\delta\left[\sqrt{-g} {\cal L}_m \left(g^{\mu\nu},g^{\mu\nu}_{\,\,\,\,\,,\lambda} \right)\right] = \frac{1}{2} \, \sqrt{-g} \, \delta g^{\mu\nu}  T_{\mu\nu}.
\end{equation}
Here, we have used the generic definition for the Energy-Momentum (EM) tensor:
$T_{\mu\nu} = 2\frac{\delta {\cal L}_m}{\delta g^{\mu\nu}}- g_{\mu\nu} {\cal L}_m$.

\subsection{Variation of the gravitational action on a curved spacetime}

Now we consider the gravitational action. Its variation is
\begin{equation}
\delta\left[\sqrt{-g} \, R\right] = \sqrt{-g} \left[
\delta g^{\alpha\beta} \, G_{\alpha\beta} + g^{\alpha\beta} \, \delta R_{\alpha\beta} \right],
\end{equation}
where $G_{\alpha\beta} = R_{\alpha\beta} - {1\over2} g_{\alpha\beta} \,R$ is the Einstein tensor and
\begin{equation}\label{lig}
g^{\alpha\beta} \, \delta R_{\alpha\beta} \equiv \nabla_{\mu} W^{\mu}.
\end{equation}
Here, the tetra-vector $W$, has components $W^{\mu}$
\begin{eqnarray}
W^{\mu} & = & \frac{1}{2} \left[ \delta g^{\lambda\nu} g^{\alpha\mu}  \left( \delta g_{\alpha\nu,\lambda} +  g_{\lambda\nu,\alpha} -  g_{\alpha\lambda,\nu} \right)\right. \nonumber \\
& + & g^{\alpha\mu} g^{\lambda\nu} \left( \delta g_{\alpha\nu,\lambda} + \delta g_{\lambda\nu,\alpha} - \delta g_{\alpha\lambda,\nu} \right) \nonumber \\
& - & \delta g^{\mu\nu} g^{\alpha\lambda} \left( g_{\alpha\nu,\lambda} + g_{\lambda\nu,\alpha} -  g_{\alpha\lambda,\nu} \right) \nonumber \\
& - &  \left.g^{\alpha\lambda} g^{\mu\nu} \left( \delta g_{\alpha\nu,\lambda} + \delta g_{\lambda\nu, \alpha} - \delta g_{\alpha\lambda, \nu} \right) \right],
\end{eqnarray}
where we have made use of the fact that $\delta g^{\mu\nu} = -g^{\mu\rho} g^{\nu\sigma}\delta g_{\rho\sigma}$.
Therefore, the first variation of the action will be
\begin{equation}\label{delta}
\delta I = \int d^4 x \sqrt{-g} \left[ \delta g^{\alpha\beta} \left( G_{\alpha\beta} + \kappa T_{\alpha\beta}\right) + g^{\alpha\beta} \delta R_{\alpha\beta} \right],
\end{equation}
with (\ref{lig})\footnote{The reader can see, for example, the page 75 in \cite*{EH}.}.

When we deal with a manifold ${\cal M}$ wich has a boundary $\partial {\cal M}$, the action should be supplemented
by a boundary term in order to the variational principle to be well-defined.
One solution of this problem was introduced by adding a term in the Hilbert-Einstein action. This additional term is known as the York-Gibbons-Hawking action\cite*{YGH,YGH1}. In this work we shall propose another solution for this problem. We shal consider In this case
\begin{equation}
\nabla_{\alpha} W^{\alpha}=\Phi(x^{\alpha}), \label{3}
\end{equation}
is nonzero. Here, $\Phi(x^{\alpha})$ is an arbitrary scalar field which becomes zero when the manifold has no boundary $\Phi=0$. In order to $\delta I=0$, in (\ref{act}), we shall
consider the condition: $ G_{\alpha\beta} + \kappa T_{\alpha\beta} = \Lambda\, g_{\alpha\beta}$, where $\Lambda$ is the cosmological constant. In this case the dynamics of the system will
be given by
\begin{eqnarray}
G_{\alpha\beta} & + & \kappa\, T_{\alpha\beta} = \Lambda\, g_{\alpha\beta}, \label{1} \\
\nabla_{\alpha} W^{\alpha} &=& \Phi, \label{2}
\end{eqnarray}
with the constriction $\delta g_{\alpha\beta} \Lambda = \Phi\, g_{\alpha\beta}$.

\subsection{Variation of the Einstein equations and Gauge-invariance}

In order to study the dynamics of the metric fluctuations we must variate the Einstein equations (\ref{1}):
\begin{equation} \label{3}
\delta G_{\alpha\beta} -\Lambda \delta g_{\alpha\beta} = - \kappa \, \delta T_{\alpha\beta},
\end{equation}
with the constrictions (\ref{2}). Using the fact that $ R=g^{\alpha\beta} R_{\alpha\beta}$ and $T= g^{\alpha\beta} T_{\alpha\beta}$, we obtain that $ \delta R = \kappa \, \delta T$,
and we obtain that the equation (\ref{3}) can be re-written as
\begin{equation}\label{4}
\delta R_{\alpha\beta} - \Lambda \delta g_{\alpha\beta} = - \kappa \,\delta S_{\alpha\beta},
\end{equation}
where we have introduced the tensor $S_{\alpha\beta} = T_{\alpha\beta} -\frac{1}{2} T \, g_{\alpha\beta}$, which takes into account matter as a source of the Ricci tensor $R_{\alpha\beta}$.
Another manner to write the equation (\ref{4}) is in terms of the tetra-vector $W^{\mu}$, is
\begin{equation}
\nabla_{\beta} W_{\alpha} - \Lambda \delta g_{\alpha\beta}  = - \kappa \, \delta{ \left( T_{\alpha\beta} - \frac{1}{2} T \,g_{\alpha\beta} \right)},
\end{equation}
which explicitly holds
\begin{eqnarray}
 \delta  g ^{\nu}_{\,\,\, \alpha , \beta\nu} &-& g^{\nu\gamma}_{\,\,\, ,\beta\nu} \delta g_{\alpha\gamma} - g^{\nu\gamma}_{\,\,\, , \alpha} \delta g_{\beta\gamma,\nu} -
g^{\nu\gamma}_{\,\,\, , \nu} \delta g_{\alpha\gamma,\beta}  + \frac{\Lambda}{2} \delta g_{\alpha\beta} \nonumber \\
&=& \frac{\kappa}{2} \left[ \delta T_{\alpha\beta} - \frac{1}{2} \, \delta \left( T \, g_{\alpha\beta} \right) \right],
\end{eqnarray}
where we have used (\ref{2}). The tetra-vector components $W_{\alpha}$ can be written as
\begin{eqnarray}
W_{\alpha} &= & \frac{1}{2} \left\{\delta g^{\gamma\theta} \left( g_{\alpha\theta,\gamma} + g_{\gamma\theta,\alpha} - g_{\alpha\gamma,\theta} \right) \right. \nonumber \\
&+& g^{\gamma\theta} \left( \delta g_{\alpha\theta,\gamma} + \delta g_{\gamma\theta,\alpha} - \delta g_{\alpha\gamma,\theta} \right)   \nonumber \\
& - & g^{\gamma\theta} \left[ \delta g_{\alpha}^{\beta} \left( g_{\gamma\beta,\theta} + g_{\theta\beta,\gamma} - g_{\gamma\theta,\beta} \right) \right.\nonumber \\
&-& \left.\left.\left( \delta g_{\gamma\alpha,\theta} + \delta g_{\theta\alpha,\gamma} - \delta g_{\gamma\theta,\alpha} \right) \right] \right\}.
\end{eqnarray}
Now, we can propose the existence of a tensor field $\Psi_{\alpha\beta}$, such that $\delta R_{\alpha\beta}\equiv \nabla_{\beta} W_{\alpha}-\Phi \,g_{\alpha\beta} \equiv \Box \Psi_{\alpha\beta} -\Phi \,g_{\alpha\beta} =- \kappa \,\delta S_{\alpha\beta}$, and hence
\begin{equation}
W_{\alpha} = \nabla^{\beta} \Psi_{\alpha\beta}.
\end{equation}
This means that the gravitational waves can be described in two manners, as the resulting of the wave equation of motion for the tensor field $\Psi_{\alpha\beta}$:
\begin{equation}\label{6}
\Box \Psi_{\alpha\beta}  -\Phi \,g_{\alpha\beta} = - \kappa \, \delta{ S_{\alpha\beta}},
\end{equation}
or as the solution of a vectorial  differential equation
\begin{equation}\label{7}
\nabla_{\beta} W_{\alpha} -\Phi \,g_{\alpha\beta} = -\kappa \,\delta{ S_{\alpha\beta}}.
\end{equation}
Notice that the field $W_{\alpha}$  and is gauge-invariant under transformations
$\bar{W}_{\alpha} = W_{\alpha} - \nabla_{\alpha} \Phi $, when the scalar function $\Phi$ complies $\Box \Phi =0$. The same is valid for $\bar{\Psi}_{\alpha\beta}
=\Psi_{\alpha\beta} - \Phi \, g_{\alpha\beta}$.

Finally, one can make the transformation:
\begin{equation}
\bar{G}_{\alpha\beta} = {G}_{\alpha\beta} - \Lambda\, g_{\alpha\beta},
\end{equation}
and the transformed Einstein equations with the equation of motion for the transformed gravitational waves hold
\begin{eqnarray}
&& \bar{G}_{\alpha\beta} = - \kappa\, {T}_{\alpha\beta}, \label{e1} \\
&& \Box \bar{\Psi}_{\alpha\beta} =- \kappa \,\delta S_{\alpha\beta}, \label{e2}
\end{eqnarray}
with $\Box \Phi=0$ and $\Phi g_{\alpha\beta} = \delta g_{\alpha\beta} \Lambda$.

\subsection{The local vacuum and dynamics of gravitational waves}

We shall introduce the following conditions
\begin{eqnarray}
&& R + 2 \kappa   {\cal L}_m =2\Lambda, \label{uno}\\
&& \delta \left(R + 2 \kappa   {\cal L}_m\right)=0, \label{dos}
\end{eqnarray}
where $\Lambda$ is the cosmological constant. Physically, the eq. (\ref{uno}) means that for each point of the space-time the action density is a constant related with the cosmological constant. The second eq. (\ref{dos}) means that its variation is also null and each alteration of the space-time is locally produced by a local variation of the physical fields of the system which we are considering.
If we use the equations (\ref{7}), we obtain
\begin{equation}
\nabla_{\beta} W_{\alpha} - \Lambda\,\, \delta g_{\alpha\beta} = -2 \kappa \,\,\delta{\left(\frac{\delta {\cal L}_m}{\delta g^{\alpha\beta}} \right)}.
\end{equation}
We define $\nabla_{\beta}\bar{W}_{\alpha} =  \nabla_{\beta}{W}_{\alpha} - \Lambda\,\, \delta g_{\alpha\beta}$, and we make $\bar{W}_{\alpha} = \nabla^{\beta} \bar{\Psi}_{\alpha\beta}$. Hence, we obtain the following wave equation for $\bar{\Psi}_{\alpha\beta}$:
\begin{equation}
\Box \bar{\Psi}_{\alpha\beta} = -2 \kappa \,\,\delta{\left(\frac{\delta {\cal L}_m}{\delta g^{\alpha\beta}} \right)}.
\end{equation}
This result is valid for an arbitrary physical system with an arbitrary Lagrangian density ${\cal L}_m$.
In the following section we shall study the dynamics of gravitational waves in the very early universe.

\section{An example: GW from the primordial Big-Bang of the universe}

As an example we can study the example which describes the gravitational waves in the primordial universe. If the expansion is driven by a scalar field $\varphi(x^{\alpha})$ which is minimally coupled to gravity
\begin{equation}
{\cal I} = \int_{V} d^4 x \sqrt{-g} \,  \left[\frac{R}{2\kappa} +\frac{1}{2} g^{\mu\nu} \varphi_{,\mu} \varphi_{,\nu} - V(\varphi)\right],
\end{equation}
where $\kappa = 8 \pi G$, $G$ is the gravitational constant, $\sqrt{-g} = a^3(t)$ is the volume of the manifold $\mathcal{M}$ and $g_{\mu\nu} = {\bf diag}[1, -a^2, -a^2, -a^2]$ are the components of the diagonal tensor metric. The dynamics of the scalar field being given by the equation
\begin{equation}
\ddot\varphi + 3 H \dot\varphi - \frac{1}{a^2} \nabla^2 \varphi + V'(\varphi)=0.
\end{equation}
Given the quantum nature of the fields $\varphi$ and $\Pi^0= {\delta {\cal L}_m \over \delta \dot\varphi}$ it seems
convenient to use the quantization procedure. To do it, we impose
the commutation relations
\begin{equation}\label{w11}
\left[\varphi(t,\vec R), \Pi^0_{(\varphi)}(t,\vec R')\right] = i\,
\delta^{(3)}\left(\vec x-\vec x'\right).
\end{equation}

\subsection{The scalar field dynamics}

We shall consider a semiclassical approach to the scalar field: $\varphi(x^{\alpha}) = \phi(t) + \delta\phi(x^{\alpha})$, such that $\phi(t)=\left<E|\varphi|E\right>$ is the background solution that describe the dynamics on the background metric. Here, $\left.|E\right>$ is some quantum state such that $\left<E|\varphi|E\right>$ denotes the expectation value of the
$\varphi$ on the 3D Euclidean hypersurface and $\phi(t)$ is the background solution of the equation
\begin{equation}\label{phi}
\ddot\phi + 3 H \dot\phi  + V'(\phi)=0.
\end{equation}
Furthermore $\delta\phi(x^{\alpha})$ are the fluctuations with respect to the background, such that $\left<E|\delta\phi|E\right>=0$. The dynamics of the fluctuations $\delta\phi$ can be approximated to
\begin{equation}
\ddot{\delta\phi} + 3 H \dot{\delta\phi} - \frac{1}{a^2} \nabla^2{\delta\phi} + V''(\phi) \, {\delta\phi}=0.
\end{equation}
Here, $V''\equiv \left.{\delta^2\,V\over \delta \varphi^2}\right|_{\phi} \,\equiv m^2$ gives the square mass of the inflaton field related to the density potential $V(\varphi)$.
From the Einstein equations, we obtain that $R+4\Lambda-\kappa T=0$, so that after making use of the fact that $T = 4 V(\phi) - \dot\phi^2$ and the condition (\ref{uno}), we obtain
\begin{eqnarray}
V(\phi) &=& \frac{3\Lambda}{\kappa} , \\ \label{ss}
\dot\phi^2 & + &\frac{6}{\kappa} \left( 2 H^2 + \dot{H} \right) = \frac{8\Lambda}{\kappa}. \label{s}
\end{eqnarray}
From the Eq. (\ref{s}) it is obvious that $V(\phi)$ is a constant, and the Eqs. (\ref{s}) with (\ref{phi}) provide us with the dynamics of
$\phi(t)$ and $H(t)$. Since $V'=0$, hence the dynamics of the background scalar field $\phi$ decouples with gravity. From the Eq. (\ref{phi}) we obtain general solution for the scalar field
\begin{equation}\label{phi1}
\dot\phi(t)=\frac{C}{a^{3}(t)}.
\end{equation}
If we replace this solution in (\ref{s}), we obtain that there are only three possible solutions.

\begin{figure*}
\includegraphics[height=10cm]{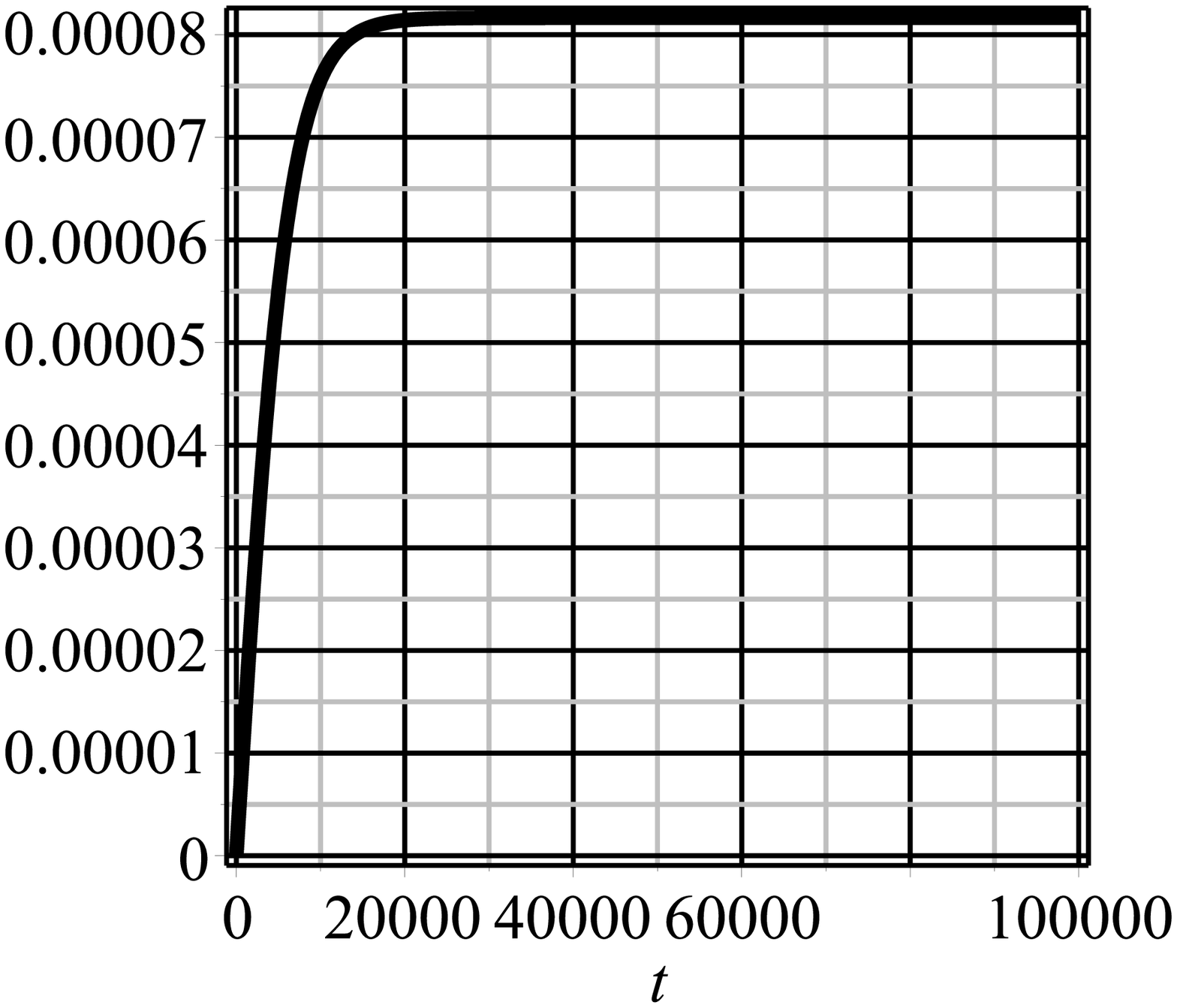}\vskip 3cm\caption{\label{f1}  Evolution of $H(t)$ during during the primordial bang, for $\Lambda=10^{-8}\, G^{-1/2}$.
The universe approaches to a asymptotic de Sitter expansion from a initial static universe. }
\end{figure*}

\begin{figure*}
\includegraphics[height=10cm]{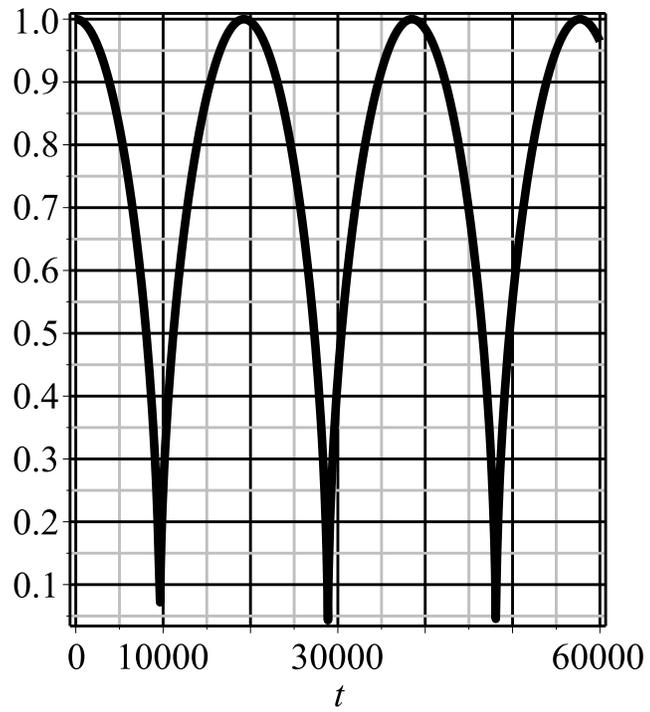}\vskip 2cm\caption{\label{f2} Evolution of the scale factor $a(t)$ for the universe with negative cosmological constant $\Lambda=-10^{-8}\, G^{-1/2}$. Notice that this universe collapses and bounce cyclically, but fails to progress.}
\end{figure*}

\begin{enumerate}

\item{The static universe}\label{stat}

The more trivial case with $C\neq 0$ and $a(t)=const.$, for $\Lambda >0$. This solution
give us a null Hubble parameter and
hence does not describe an expanding universe.

\item{The bang of the primordial universe (or pre-inflation)}\label{preinf}

The more interesting case is with $C=0$ (i.e., with $\dot{\phi}=0$), for $\Lambda >0$. The dynamics of the Hubble parameter is given by the equation
$12 H^2 + 6\dot{H}=8\Lambda$, and the solution is
\begin{equation}\label{h}
H(t) = \sqrt{\frac{2\Lambda}{3}} \, \tanh{\left[2\sqrt{\frac{2\Lambda}{3}}\, t\right]},
\end{equation}
which is related to a scale factor
\begin{equation}\label{aa}
a(t)= \frac{a_0}{\left[1-\tanh^2{\left(2\sqrt{\frac{2\Lambda}{3}}\, t\right)}\right]^{1/4}},
\end{equation}
with $a_0 =\sqrt{\frac{3}{2\Lambda}}$.
This interesting case presents a new paradigm in cosmology because describes an universe with a Hubble parameter that increases from a null value to an asymptotically constant value $\left.H(t)\right|_{t\gg G^{1/2}} \rightarrow\sqrt{\frac{2}{3} \Lambda}$, describing the creation of the universe and its transition from a static state {\bf\ref{stat}},
to an accelerated de Sitter inflationary expansion. The Hubble parameter was plotted in the figure (\ref{f1}). A special case of this case with $\dot{H}=0$ describes a de Sitter inflationary expansion governed by the cosmological constant $\Lambda >0$ with a scale factor $a(t)/a_0=e^{\sqrt{\frac{2\Lambda}{3}} t}$ and $C=0$. This is the asymptotic solution of (\ref{aa}), for very large times.

\item{The oscillating universe}\label{oscill}

To finalize there is a case with $C=0$ and $\Lambda <0$ in which the solution of $a(t)$ is oscillating [see figure (\ref{f2})]
\begin{equation}
a(t)= \frac{a_0}{\left[\sec^2{\left(2\sqrt{\frac{-2\Lambda}{3}}\, t\right)}\right]^{1/4}}.
\end{equation}
The Hubble parameter for this case is
\begin{equation}
H(t)= -\sqrt{\frac{-2\Lambda}{3}} \, \tan{\left[2\sqrt{\frac{-2\Lambda}{3}}\, t\right]}.
\end{equation}
This interesting case describes an universe that fails to progress. It could be assimilated to a bouncing universe\cite*{bounce,bounce1}.
\end{enumerate}

\subsection{GW from the primordial bang of the universe}

We shall study GW for the case {\bf\ref{preinf}}, which are the more interesting. Due to the fact that $\frac{\delta {\cal L}_m}{\delta{g^{\alpha\beta}}} = \varphi_{,\alpha} \varphi_{,\beta}$, the linearized variation $ \delta{\left(\frac{\delta {\cal L}_m}{\delta{g^{\alpha\beta}}}\right)}= \varphi_{,\alpha} \,\varphi_{,\beta} - \phi_{,\alpha} \, \phi_{,\beta}$, will be
\begin{equation}
\delta{\left(\frac{\delta {\cal L}_m}{\delta{g^{\alpha\beta}}}\right)} =  \phi_{,\alpha}\, \delta\phi_{,\beta} ,
\end{equation}
so that
\begin{equation}
\Box \bar{\Psi}_{\alpha\beta} = - 2 \kappa \, \phi_{,\alpha}\, \delta\phi_{,\beta}.
\end{equation}
Using (\ref{phi1}) we obtain that $\phi_{,\alpha}=0$, $\forall \alpha$, so that finally we obtain the equation of motion for $\Psi_{\alpha\beta}$
\begin{equation}
\Box \bar{\Psi}_{\alpha\beta} = 0,
\end{equation}
which means that $g_{\alpha\beta} = \Lambda \delta g_{\alpha\beta}$.

\subsection{Transverse-Traceless (TT) Gauge}

The expansion of the tensor field $\bar{\Psi}_{ab}(t,\vec{x})$ can be made as
\begin{eqnarray}\label{gww}
\bar{\Psi}_{ab}(t,\vec{x}) &=& \frac{1}{(2\pi)^{3/2}}\,\sum_{A=+,\times} \,\int d^3 k \,\, e^A_{ab}(\hat{z})\nonumber \\
&\times & \left[
B_k \, e^{i\vec{k}.\hat{z}}  \, \xi_c(t) + B^{\dagger}_k \, e^{-i\vec{k}.\hat{z}}  \, \xi^*_c(t)\right],
\end{eqnarray}
where $a,b=1,2$, denote the transverse polarizations $+,\times$, on the plane with normal co-linear with $\vec{k}$ and $e^A_{ab}$ are the components of the polarization tensor, such that
$e^A_{ab} \, \bar{e}^{ab}_{A'} = \delta^A_{A'}$. In the frame where $\vec{k}$ is along the $\hat{z}$ direction these polarizations are
\begin{equation}
e^+_{ab}=\left(\begin{array}{ll} 1 & \,\,\,0 \\ 0 & -1 \end{array}\right)_{ab}, \qquad \qquad e^{\times}_{ab}=\left(\begin{array}{ll} 0 & 1 \\ 1 & 0 \end{array}\right)_{ab},
\end{equation}
with $a,b$ spanning the $(x,y)$ plane.

In order to solve the equations for the gravitational waves we shall use the TT gauge, which is represented by the following conditions
\begin{equation}
\bar{\Psi}_{0\mu}=0, \qquad \bar{\Psi}^i_{\,\, i} =0, \qquad \nabla^j \bar{\Psi}_{ij}=0.
\end{equation}
The equation of motion for the modes $\xi_k(t)$ is
\begin{equation}\label{xi}
\ddot\xi_c(t) + 3 \frac{\dot a}{a} \dot\xi_c(t) + \frac{k^2}{a(t)^2} \xi_c(t)=0.
\end{equation}
The annihilation and creation operators $B_{k}$ and
$B_{k}^{\dagger}$ satisfy the usual commutation algebra
\begin{equation}\label{m5}
\left[B_{k},B_{k'}^{\dagger}\right]=\delta ^{(3)}(\vec{k}-\vec{k'}),\quad
\left[B_{k},B_{k'}\right]=\left[B_{k}^{\dagger},B_{k'}^{\dagger}\right]=0.
\end{equation}
Using the commutation relation (\ref{w11}) and the Fourier
expansions (\ref{gww}), we obtain the normalization condition for
the modes. For convenience we shall re-define the dimensionless time: $\tau =b\,t$, where $b=\sqrt{\frac{2\Lambda}{3}}=\frac{1}{a_0}$, so that the normalization condition for  $\xi_{c}(\tau)$ is
\begin{equation}\label{m6}
\xi_{c}(t) \frac{d{\xi}^*_{c}(\tau)}{d\tau} - \xi^*_{c}(t) \frac{{\xi}_{c}(\tau)}{d\tau} = i
\,\left(\frac{a_0}{a(\tau)}\right)^{3},
\end{equation}
where the asterisk denotes the complex conjugated.
For the case {\bf\ref{preinf}}, in which the Hubble parameter and the scale factor are given respectively by (\ref{h}) and (\ref{aa}), the general solution for the amplitudes
$\xi_c(\tau)$ is
\begin{eqnarray}
\xi_c(\tau) &=& C_1\, \frac{\sinh{(\tau)}}{\sqrt{2\cosh^2{(\tau)}-1}} \nonumber \\
 &\times &{\rm Hn}\left[-1,\frac{c^2-1}{4};0,\frac{1}{2},\frac{3}{2},\frac{1}{2};-\tanh^2{(\tau)}\right] \nonumber \\
&+& C_2\, \frac{\cosh{(\tau)}}{\sqrt{2\cosh^2{(\tau)}-1}} \nonumber \\
&\times & {\rm Hn}\left[-1,\frac{c^2+1}{4};-\frac{1}{2},0,\frac{1}{2},\frac{1}{2};-\tanh^2{(\tau)}\right] , \nonumber \\
\end{eqnarray}
where ${\rm Hn}[a,q;\alpha,\beta,\gamma,\delta;z]=\sum^{\infty}_{j=0} c_j\, z^j$ is the Heun function. Since the Heun functions are written as infinity series, we can make a series expansion in the both sides of (\ref{m6}), in order to obtain the restrictions for the coefficients $C_1$ and $C_2$, and the wavenumber values $k$. From the zeroth order expansion (in $\tau$), we obtain
that $C_2=i\,C_1/2$. Hence, we shall choose $C_1=1$ results that $C_2=i/2$\footnote{The polynomial expansion of $\xi_{c}(t) \frac{d{\xi}^*_{c}(\tau)}{d\tau} - \xi^*_{c}(t) \frac{{\xi}_{c}(\tau)}{d\tau} = i
\,\left(\frac{a_0}{a(\tau)}\right)^{3}$, is given by
\begin{equation}
\xi_{c}(t) \frac{d{\xi}^*_{c}(\tau)}{d\tau} - \xi^*_{c}(t) \frac{{\xi}_{c}(\tau)}{d\tau} - i\,\left(\frac{a_0}{a(\tau)}\right)^{3} = \sum^{\infty}_{N=1} f_N(c)\, \tau^N=0,
\end{equation}
where $f_N(c^{(N)}_n)=0$, for each $N$. There are $2N$ modes for each $N$-th order of the expansion. }.

Of course the series is infinite so that there are infinite values of quasi-normal modes which has zero norm on the plane orthogonal to the direction of propagation:$\hat{z}$: $\vec{k} =c^{(N)}_n \hat{e^1} + (\pm i c^{(N)}_n) \hat{e^2} + k \hat{e^3}$, with complex values
$c^{(N)}_n$ which provide us the quantization of tensor fields $\bar\Psi_{ab}$. Notice that $\|\vec{k}\|^2=k^2$, so that the norm of the polarization vectors on the plane ($\hat{e^1},\hat{e^2}$), is zero.
When the polarization modes which are purely imaginary or purely reals correspond to values with polarization $+$. On the other hand, modes with complex-$c^{(N)}_n$ correspond to those with polarization $\times$.

We can calculate the two-point expectation value for the fluctuations of the spacetime due to gravitational waves at the spatial points $\vec{x}$. If the spatial position in
the interior of the exploiting source is denoted by $\vec{x}'$, we have
\begin{eqnarray}
&&\left<E |\bar{\Psi}^2|E\right> (\tau,\vec{x},\vec{x}') \nonumber \\
&=& 2\,i\,\sum_{A=+,\times} \sum^{\infty}_{N=1} \sum^{2N}_{n=1} \sin{\left[\vec{c}^{(N)}_n . \left(\vec{x} - \vec{x}'\right)\right]} \,\xi_{c^{(N)}_n}(\tau) \xi^*_{c^{(N)}_n}(\tau). \nonumber \\
\end{eqnarray}
Hence, integrating on all the points of the spherical source of ratio $b^{-1}$, we obtain
\begin{eqnarray}
\left<E|\bar{\Psi}^2|E\right>(\tau,\vec{x},\theta) & = & 8\pi\,i\,\sum_{A=+,\times} \sum^{\infty}_{N=1} \sum^{2N}_{n=1}   \nonumber \\
& \times & \,\xi_{c^{(N)}_n}(\tau) \xi^*_{c^{(N)}_n}(\tau) \,I^{(N,n)}(|\vec{x}|,\theta), \nonumber \\
\end{eqnarray}
where $0<\theta < \pi/2$ is the angle between $\vec{x}$ and $\vec{x}'$, and the function $I^{(N,n)}(|\vec{x}|,\theta)$ for the $N$-th order in the expansion is given by
\begin{small}
\begin{eqnarray}
&&I^{(N,n)}(|\vec{x}|,\theta) = \left[ \left(c^{(N)}_n\right)^3\, \cos^3{\left(\theta\right)}\right]^{-1}\nonumber \\
&\times &  \left\{\cos{\left[c^{(N)}_n \left(|\vec{x}| - \frac{\cos(\theta)}{b}\right)\right]} \left(\frac{\cos(\theta)}{b}\right)^2 \left(c^{(N)}_n\right)^2 \right.\nonumber \\
&+& 2 \cos{\left(c^{(N)}_n |\vec{x}|\right)} + 2 \sin{\left[c^{(N)}_n \left(|\vec{x}| - \frac{\cos(\theta)}{b}\right)
\right]}\left(\frac{\cos(\theta)}{b}\right) c^{(N)}_n \nonumber \\
&-& \left. 2\cos{\left[c^{(N)}_n \left(|\vec{x}| - \frac{\cos(\theta)}{b}\right)\right]}\right\}, \nonumber \\
\end{eqnarray}
\end{small}
for $|\vec{x}| > 1/b =\sqrt{\frac{3}{2\Lambda}}$.
Finally, we can calculate the energy density due to gravitational waves, $\rho_{{gw}}=\left<E|\dot{\bar{\Psi}}^2|E\right>(t,\vec{x},\theta)$
\begin{eqnarray}
&& \rho_{{gw}}(\tau,\vec{x},\theta) = \nonumber \\
&& 8\pi\,i\,\sum_{A=+,\times} \sum^{\infty}_{N=1} \sum^{2N}_{n=1} \,\dot\xi_{c^{(N)}_n}(\tau) \dot\xi^*_{c^{(N)}_n}(\tau) \,I^{(N,n)}(|\vec{x}|,\theta), \nonumber \\
\end{eqnarray}
which up to fifth order in the series expansion with respect to $\tau$, is
\begin{eqnarray}
&&\rho_{{gw}}(\tau,\vec{x},\theta) = \nonumber \\
&& 8\pi\,i\,\sum_{A=+,\times} \sum^{\infty}_{N=1} \sum^{2N}_{n=1} \left[1+ \sum^{2N}_{n=1} F^{(N)}_n(c^{(N)}_n) \,\tau^N \, I^{(N,n)}(|\vec{x}|,\theta)\right], \nonumber \\
\end{eqnarray}
where the coefficients $F^{(N)}_n(c^{(N)}_n)$, are
\begin{eqnarray}
&& F^{(1)}_n( c^{(1)}_n ) = \frac{1}{3}\left[1-\left(c^{(1)}_n\right)^2\right], \\
&& F^{(2)}_n(c^{(2)}_n)  =  \frac{1}{9} \left[\frac{53}{20} \left(c^{(2)}_n\right)^4+ \frac{23}{5} \left(c^{(2)}_n\right)^2-50\right], \\
&& F^{(3)}_n(c^{(3)}_n)=  -\left[\frac{109}{1260} \left(c^{(3)}_n\right)^6- \frac{37}{140} \left(c^{(3)}_n\right)^4 \right. \nonumber \\
&& \left.- \frac{1693}{630} \left(c^{(3)}_n\right)^2+ \frac{128}{63}\right], \\
&& F^{(4)}_n(c^{(4)}_n)  =  \left[\frac{319}{28350} \left(c^{(4)}_n\right)^8-\frac{6281}{56700} \left(c^{(4)}_n\right)^6 \right. \nonumber \\
&& \left.-\frac{49877}{56700} \left(c^{(4)}_n\right)^4-\frac{4807}{5670} \left(c^{(4)}_n\right)^2+\frac{527}{28}\right], \\
&& F^{(5)}_n(c^{(5)}_n)  =  -\left[\frac{1493}{1871100} \left(c^{(5)}_n\right)^{10}- \frac{6943}{467775} \left(c^{(5)}_n\right)^8 \right. \nonumber \\
&& \left. - \frac{351}{1925} \left(c^{(5)}_n\right)^6+ \frac{1839031}{1871100} \left(c^{(5)}_n\right)^4 \right.\nonumber \\
&+& \left.\frac{463679}{53460} \left(c^{(5)}_n\right)^2- \frac{63541}{8316}\right].
\end{eqnarray}
In the table we have included the square wavenumbers for the first five orders of the expansion.
\begin{small}
\begin{center}
\begin{tabular}{| r | l  r | l | r | l | r |}
\hline
\small Square wavenumbers $c^{(N)}_n$  & \small $\left(c^{(N)}_n\right)^2 $ values    \\ \hline
\small $\left(c^{(1)}\right)^2$ & \small -0.5   \\  \hline
$\left(c^{(2)}\right)^2$ & \small 6.57  & \small -0.57   \\ \hline
\small $\left(c^{(3)}\right)^2$ & \small -53.78  & \small 0.01 + 1.25 i & \small 0.01 - 1.25 i \\ \hline
\small $\left(c^{(4)}\right)^2$ & \small 9.60 & \small -1.29 & \small 9.85 + 9.40 i  & \small 9.85 - 9.40 i  \\ \hline
\small $\left(c^{(5)}\right)^2$ & \small 5.93 & \small 12.72 & \small -1.34 & \small 13.84 + 18.30 i & \small 13.84 - 18.30 i  \\ \hline
\end{tabular}\label{tt}
\end{center}
\end{small}

\section{Final Comments}

We have studied a new approach to gravitational waves in which the Einstein equations are governed by the cosmological constant. This constant appears related to
the existence of a manifold which is closed. Under these circumstances the Einstein-Hilbert action remains invariant when the
gravitational are transformed as: $\bar{\Psi}_{\alpha\beta} =  {\Psi}_{\alpha\beta} - \Phi g_{\alpha\beta}$, with $\delta  \Lambda = \Phi g_{\alpha\beta}$.
A simple example where the matter Lagrangian is described by a scalar field is studied an the resulting dynamics is surprising. There are only three
dynamical solutions. i) The first one describe that a primordial universe remains static. ii) The second one describe a primordial universe which is initially static but
begins to increase until an inflationary stage. iii) The third solution (governed by a negative cosmological solution) describes an eternally oscillating (bouncing) universe from the initial state.

We have explored the study of gravitational waves for the case ii), because describes a pre-inflationary universe that evolves towards an asymptotic inflationary phase. Finally, we have calculated the dynamics of the GW in this stage.
The normalization conditions for the modes impose that the wavenumbers for $\bar{\Psi}_{\alpha\beta}$ are an infinity number of values for the polarization modes, which can be real and imaginary for the $+$-modes and complex for the $\times$-modes.

 \section*{Acknowledgements}

\noindent L. S. Ridao and M. Bellini acknowledge
UNMdP and  CONICET (Argentina) for financial support.

\newpage

\newpage

\end{document}